\documentclass{pas}

\usepackage{multirow}
\usepackage{graphicx}

\newcommand{\rsun}{R$_{\odot}$}			
\newcommand{\lsun}{L$_{\odot}$}			
\newcommand{\rjup}{R$_{\rm Jup}$}		
\newcommand{\msun}{M$_{\odot}$}			
\newcommand{\mjup}{M$_{\rm Jup}$}		
\newcommand{\one}{51 Eridani}	

\begin{document}
	
	\lefttitle{Measuring the Stellar and Planetary Parameters of the 51 Eridani System}
	\righttitle{A. Elliott et al.}
	
	\jnlPage{1}{10}
	\jnlDoiYr{2023}
	\doival{10.1017/pasa.xxxx.xx}
	
	
	\title{Measuring the Stellar and Planetary Parameters of the 51 Eridani System}
	
	\author{\sn{Ashley} \gn{Elliott}$^{1}$, \sn{Tabetha} \gn{Boyajian}$^{1}$, \sn{Tyler} \gn{Ellis}$^{1}$, \sn{Kaspar} \gn{von Braun}$^{2}$, \sn{Andrew} \gn{W. Mann}$^{3}$, and \sn{Gail} \gn{Schaefer}$^{4}$}

	\affil{$^1$Department of Physics and Astronomy, Louisiana State University, Baton Rouge, LA 70802, USA\\
		$^2$ Lowell Observatory, 1400 West Mars Hill Rd. Flagstaff, AZ, 86001, USA\\
		$^3$ Department of Physics and Astronomy, The University of North Carolina at Chapel Hill, Chapel Hill, NC 27599, USA \\
		$^4$ The CHARA Array of Georgia State University, Mount Wilson Observatory, Mount Wilson, CA 91023, USA \\}
	
	\corresp{A. Elliott, Email: aelli76@lsu.edu}
	
	
	\history{(Received xx xx xxxx; revised xx xx xxxx; accepted xx xx xxxx)}

	\begin{abstract}
		In order to study exoplanets, a comprehensive characterization of the fundamental properties of the host stars - such as angular diameter, temperature, luminosity, and age, is essential, as the formation and evolution of exoplanets are directly influenced by the host stars at various points in time. In this paper, we present interferometric observations taken of directly imaged planet host \one~at the CHARA Array. We measure the limb-darkened angular diameter of \one~to be $\theta_{\rm LD} = 0.450\pm 0.006$~mas and combining with the Gaia zero-point corrected parallax, we get a stellar radius of $1.45 \pm 0.02$~\rsun. We use the PARSEC isochrones to estimate an age of $23.2^{+1.7}_{-2.0}$~Myr and a mass of $1.550^{+0.006}_{-0.005}$~\msun. The age and mass agree well with values in the literature, determined through a variety of methods ranging from dynamical age trace-backs to lithium depletion boundary methods. We derive a mass of $4.1\pm0.4$~\mjup~for 51 Eri b using the Sonora Bobcat models, which further supports the possibility of 51 Eri b forming under either the hot-start formation model or the warm-start formation model.
	\end{abstract}
	
	
	\maketitle
	
	\section{Introduction}
	
	Young, nearby stars are of major interest to scientists as they can provide insight into stellar evolution, as well as planetary formation. The exoplanet host \one~has been an object of interest in the astronomical community for a number of years. \one~is a F0 spectral type, $V = 5.22$ mag star and has a binary pair companion, GJ 3305A,B (both M0 dwarfs). \one~and GJ 3305A,B have a separation of 66$^{\prime\prime}$, or roughly 2000~au, and the system components are co-moving together in a hierarchical relationship \citep{fieg06}. The exoplanet, 51 Eri b, has a semi-major axis of $12^{+4}_{-2}$~au, an inclination of $133^{+14}_{-7}$~deg, an eccentricity of $0.45^{+0.10}_{-0.15}$, and an orbital period of $32^{+17}_{-9}$~years, all determined by fits of SPHERE and GPI data by \cite{Maire2019}.
	
	\one~is a member of the $\beta$ Pictoris moving group ($\beta PMG$), which is one of the youngest and closest moving groups to Earth. \citet{zuc01} determined that the $\beta$ Pictoris Moving Group  consisted of 17 star systems. Since then, there are now a few hundred candidate systems belonging to the $\beta PMG$ \citep{Mir20}. The age estimates for the $\beta PMG$ include $23\pm3$ Myr \citep{mama14} using the lithium depletion boundary method and isochronal ages for FGMK stars, which is consistent with the age estimate of \one, $20 \pm 6$ Myr \citep{maci15} using the group's lithium depletion boundary age. A more recent dynamical age estimate developed by \cite{Mir20} presents an age of $18.5^{+2.0}_{-2.4}$ Myr. This estimate is a dynamical trace-back age that reconciles other trace-back estimates with other methods such as lithium depletion, isochronal ages, and other dynamical estimates.

	\citet{maci15} presented the discovery of the directly imaged planet 51 Eri b by the Gemini Planet Imager (GPI). Follow-up work by GPI and NIRC2 at the W. M. Keck 2 telescope \citep{wiz00} have confirmed and refined the orbit and planetary properties of the system. The most current study of this system by \cite{dup22} determined an upper limit for the planet mass of $\geq 10.9$ \mjup~within a $2 \sigma$ interval and an orbital separation of $10.4_{-1.1}^{+0.8}$~au derived using joint fitting of the \textit{Hipparcos-Gaia} Catalog of Accelerations (HGCA-EDR3) proper motions and relative astrometry. \cite{Sam2017} determined a planetary effective temperature of $760\pm20$ K, and a radius of $1.1^{+0.2}_{-0.1}$ \rjup~from fitting spectro-photometry from VLT/SPHERE. \cite{Brown2023} recently published updated parameters on 51 Eri b using observations from VLT/SPHERE. A radiative transfer model fit using {\fontfamily{qcr}\selectfont petitRADTRANS} resulted in a planetary effective temperature of $807 \pm 45 $~K along with a radius of $0.93 \pm 0.04$~\rjup, and a mass of $3.9 \pm 0.4$ \mjup. For the mass of the planet, \cite{Brown2023} reported three possible masses: $3.9\pm 0.4$ \mjup~as the nominal result, $2.4$ \mjup~as a mass using evolutionary models by \cite{Baraffemodels} with an age of 10 Myr, and $2.6$ \mjup~with an age of 20 Myr. These ages were taken from age estimates published by \citet{Lee2022} and \citet{maci15}, respectively. 
	
	Given the young age of the system (both star and planet) determined by several literature sources, 51 Eri b is a perfect candidate to study a planet that is both young and still being influenced by its initial conditions of formation \citep{maci15}. There are two main planet formation scenarios that most planets discovered fit into: the cold-start model and the hot-start model. The cold-start model is described by core accretion and usually results in lower entropy and a smaller radius of the planet. The hot-start model is described by disk instability which results in a planet with a higher entropy, a higher effective temperature, and a larger radius \citep{Speg12}. A third formation scenario, cleverly named the warm-start model, involves a combination of the hot and cold start formation criteria \citep{Speigel2013}. The warm-start model is a spectrum of initial conditions motivated by the observed preference of core accretion with gas giants forming closer to their stars and the recent observations of young exoplanets that are hotter than the cold-start model predictions but colder than the hot-start model predictions \citep{Speigel2013}. A planet's luminosity can provide insights into its formation because it is a function of age, mass, and initial conditions \citep{Marley07, Speg12}. \citet{maci15}'s initial observations and study of 51 Eri b noted that the core-accretion theory explains the formation of this planet due to the derived low luminosity range (log($L/\rm L_{\odot}) = -5.4~\rm to -5.8$). \cite{Sam2017} explored all three scenarios and ruled out the cold-start models due to the new luminosity ranges (log($L/\rm L_{\odot}) = -5.4~\rm to -5.5$) derived and found that the planet mass favored the hot- or warm- start models.  \cite{dup22} also ruled out the cold-start formation with their derivation of a lower limit on the initial specific entropy. Further study of this planet and its star can continue to explain its formation.

	Characterizing exoplanet host stars is key to understanding the exoplanet itself. The directly measured fundamental properties of a star, i.e. radius, effective temperature, and luminosity, lead to more precise characterization of the exoplanet. The habitable zone of a planet is heavily influenced by its host star, so in order to constrain the habitable zone boundaries and planets' effective temperature, one must know the host star's fundamental properties \citep{vonBraun17}. The derived age of a star from the directly measured fundamental properties can provide information about the formation of the exoplanet. 
	
	There are several indirect methods to determine stellar fundamental properties, such as atmospheric modeling and stellar evolutionary models. These models have been shown to have difficulties reproducing observations \citep{Boyajian12, Boyajian13}. Long baseline optical/infrared interferometry provides high angular resolution measurements of stars, allowing astronomers to calculate fundamental properties without relying on models. 
	
	Interferometric observations allow us to measure the stellar angular diameter, which when combined with the parallax, gives a direct measurement of the stellar radii, one of the fundamental parameters of an astronomical object. Optical interferometry achieves a higher resolution (on the order of milli-arcseconds) than most large telescopes by combining light from several pairs of telescopes that are separated across a variety of baselines, or the separation between telescopes. The resolution capability of an interferometer increases the precision of an angular diameter measurement which then lowers the uncertainty of stellar parameters that can be derived from the angular diameter, i.e. stellar effective temperature and linear radius.

	Similar to \one~and its planet, several other systems of stars with directly imaged exoplanets have been characterized with the CHARA Array. One such system is the $\kappa$ Andromedae and its exoplanet, $\kappa$ And b by \cite{JJKappa}. \cite{JJKappa} took into account the oblate nature and gravity darkening caused by $\kappa$~And's rapid rotation through the use of modeling interferometric observations. The model results were used to determine fundamental stellar parameters, such as temperatures at the poles and equator, surface gravities, luminosity, stellar age and mass, and planetary age and mass. 
	
	\cite{BainesE} used high resolution interferometric observations to study HR 8799, which hosts 4 directly imaged companions \citep{HR8799disc}. The classification of the 4 companions, which is highly dependent on the age of the planet (inferred from the age of the star), is the subject of debate amongst astronomers (see Table 1 in \citealt{BainesE}). \cite{BainesE} combined the angular diameter, parallax, and photometry to determine stellar parameters, such as linear radius, stellar effective temperature, and luminosity. The effective temperature calculated revealed two possible age scenarios of HR 8799: either the star is contracting onto the zero-age main sequence (ZAMS) or expanding away from it. \cite{BainesE} used the Yonsei-Yale evolutionary models to investigate both possibilities. The resulting young ages (less than 0.1 Gyr) from either scenario highly favored the classification of planet for the 4 companions, not brown dwarf. Additional recent works that use interferometric radii to determine planet properties are as follows but not limited to: \cite{Caballero2022}, \cite{Ellis2021}, and \cite{Ligi2019} with more references tabulated in the \cite{vonBraunBoya} compilation. Other works demonstrating the wide utility of interferometric data are \cite{Korolik2023} and \cite{Roettenbacher2022}, which demonstrate analyzing stellar activity and co-alignment of systems and \cite{Ibrahim2023} which studies the inner au disk of Herbig Be star HD~190073. 
	
	In this paper, we present new stellar and planetary properties of the \one~system. This paper is organized as follows: the interferometric observations in Section~\ref{s:data}, the directly determined stellar properties in Section~\ref{S:direct}, the derived stellar and planetary properties in Section~\ref{s:empire}, and a discussion of the results presented in this paper in comparison with the results published in the literature in Section~\ref{s:Disc}. 
	
	\section{Interferometric Observations}\label{s:data}
	The Center for High Angular Resolution Astronomy (CHARA) is a long baseline, optical/IR interferometer located at Mt. Wilson, CA. The CHARA Array \citep{tenB05} comprises of six 1-meter telescopes arranged in a Y-shape configuration, with three arms: East (E), West (W), and South (S), each having two telescopes labeled 1 (outermost) and 2 (innermost). Baselines for the interferometer are formed by pairs of telescopes, such as E1/W1 (outermost East and West telescopes). We observed \one~in 2015 and 2016 using the Precision Astronomical Visible Observations (PAVO) beam combiner \citep{Irel08}, which measures interference fringes over a 630-950 nm dispersed bandwidth with 2 telescopes (as indicated in Table~\ref{tab:Obs}). We observed \one~in 2021 using the CLASSIC \citep{CLASSIC} beam combiner, which operates in the near-infrared $H$-band.
	
	An observation sequence consists of the target star, or science star, and a selection of calibrator stars. Observations of calibrator stars allow for the removal of atmospheric and instrumentation noise from the observations of the science star. Data were taken in the following pattern: calibrator star - science star - calibrator star. This pattern is referred to as a bracket. The observations taken adhere to a minimum 2 night, 2 calibrator, 2 baseline requirement to reduce and/or eliminate unknown systematic errors in the visibility data. 
	
	We use the Jean-Marie Mariotti Center Stellar Diameter Catalog (JSDC) (\cite{JSDC}, \cite{JMMC})\footnote{Available at https://www.jmmc.fr/english/tools/data-bases/jsdc-72/} to find suitable calibrator stars for our science target. The calibrator stars are chosen such that they are unresolved, nearby to the science star on the sky ($<10^{\circ}$), and have no known companions or rapid rotation. A summary of observations can be found in Table~\ref{tab:Obs} and a list of calibrator stars used can be found in Table~\ref{tab:Cals}. 
	
	We reduce and calibrate data for each night of PAVO observations using the PAVO software available through CHARA's Remote Data Reduction Machine \citep{Irel08}. In the calibration process, we found that that the calibrator star, HD 29335, showed spurious results indicating that it was a bad calibrator. The amount of usable observations diminished after the removal of this calibrator from the observing sequence (seen in the parentheses in the column ``Brackets'' in Table~\ref{tab:Obs}).
	
	We reduce and calibrate data for each night of the CLASSIC observations using the CLASSIC/CLIMB reduction software {\fontfamily{qcr}\selectfont redfluor}, accessed through CHARA's Remote Data Reduction Machine (\citealp{CLASSICmath}, \citealp{CLASSICsoft}). For the first two nights of observations, only one observation of the science star and one observation of a calibrator star was made. We found that the calibrator was more resolved than the science star. Due to this trend, we suspect that this calibrator (HD 26912) is a bad calibrator and did not use in our analysis.

	\begin{table*}
		\centering
		\caption{Summary of Observations: The date of observation in UT time is listed in the first column on the left. The wavelength of observation is shown in the second column. The baseline of observation is in the third column. The number of brackets is listed in the fourth column. The number in parentheses is the actual number of brackets used for data analysis due to the bad calibrators HD 29335 and HD 26912 (marked with a $^{*}$ in the table). Finally, the calibrators used for each night are listed in the right-most column.}
		\begin{tabular}{lccccr}
			\hline
			Date [UT] & Wavelength [$\mu$m] & Baseline & Brackets & Calibrators\\
			\hline
			2015-10-11 & 0.630-0.950 & E1/W1 & 7 (5) & HD28375, HD27563, HD29335$^{*}$ \\
			2015-10-12 & 0.630-0.950  & S2/E2 & 9 (7) & HD28375, HD27563, HD29335$^{*}$ \\
			2015-11-06 & 0.630-0.950  & E2/W1 & 3 (1) & HD27563, HD29335$^{*}$ \\
			2016-11-09 & 0.630-0.950  & W2/E2 & 3 (1) & HD27563, HD29335$^{*}$ \\ 
			2021-08-25 & 1.6731  & S1/E1 & 1 (0) & HD26912$^{*}$       \\
			2021-08-26 & 1.6731  & S1/E1 & 1 (0) & HD26912$^{*}$       \\
			2021-08-27 & 1.6731  & S1/E1 & 4 (3) & HD29248, HD28736       \\ 
			\hline
		\end{tabular}
		\label{tab:Obs}
	\end{table*}

	\begin{table}
		\centering
		\caption{Summary of calibrator stars: The $R$ magnitudes and $\theta_{\rm LD}$ are taken from the JMMC Stellar Diameters Catalog v2 \citep{JSDC}. HD 29335 and HD 26912 are bad calibrators (marked with a $^{*}$).}
		\begin{tabular}{lcr}
			\hline
			Calibrator & $R$ mag & $\theta_{\rm LD}$ [mas]\\
			\hline
			HD 28375       & 5.531  & $0.179\pm0.006$ \\
			HD 27563       & 5.891  & $0.170\pm0.006$\\
			HD 29335$^{*}$ & 5.345  & $0.225\pm0.006$\\
			HD 26912$^{*}$ & 4.250  & $0.287\pm0.026$\\
			HD 29248       & 8.687  & $0.111\pm0.003$\\
			HD 28736       & 5.979  & $0.348\pm0.008$\\
			\hline
		\end{tabular}
		\label{tab:Cals}
	\end{table}
	
	\section{Directly Determined Stellar Properties}\label{S:direct}
	
	An interferometer produces interference fringes that allow us to measure two basic pieces of information: the amplitude of the fringes and the phase shift of the peaks. The amplitudes allow us to measure visibilities, which describe the fringe contrast of the interference pattern. The visibility of a star can tell us about the star's size, shape, and any surface features. Normalized visibilities are measured between 0 and 1, with a visibility of 0 indicating a completely resolved star and a visibility of 1 indicating a completely unresolved star. 
	
	When observing a star, it is important to remember that a star's brightness is not uniform across the star. Optical depth and effective temperature gradients across the disk of a star combine to create an effect called limb darkening. To an observer, a star is brighter in the center and then as you move out from the center, the brightness goes down until it is 0 at the apparent edge of the star. A star's uniform disk diameter does not take into account the limb darkening effect while the limb darkened diameter does.

	In order to determine the angular diameter of a star, the visibility squared as a function of baseline ($B$), angular diameter (either uniform, $\theta_{\rm UD}$, or limb darkened, $\theta_{\rm LD}$), and wavelength ($\lambda$) is used to fit the interferometric visibilities using Equation \ref{eq:vsquare}. 
	\begin{multline} \label{eq:vsquare}
		V^{2} = \Biggl[\left(\frac{1-\mu}{2} + \frac{\mu}{3}\right)^{-1}\cdot\left((1-\mu) \cdot \frac{J_{1}(x)}{x}\right) \\
		+ \left(\mu \sqrt{\frac{\pi}{2}} \cdot \frac{J_{3/2}(x)}{x^{3/2}}\right)\Biggr]^{2}
	\end{multline}
	
	\noindent where $x = \frac{\pi B \theta}{\lambda}$ and $\mu$ is the limb-darkening coefficient (LDC). The first iteration of this fit is applied using a $\mu$ value of 0, which corresponds to a uniform disk diameter, $\theta_{\rm UD}$. 
	
\begin{figure}
		\includegraphics[width=0.95\columnwidth]{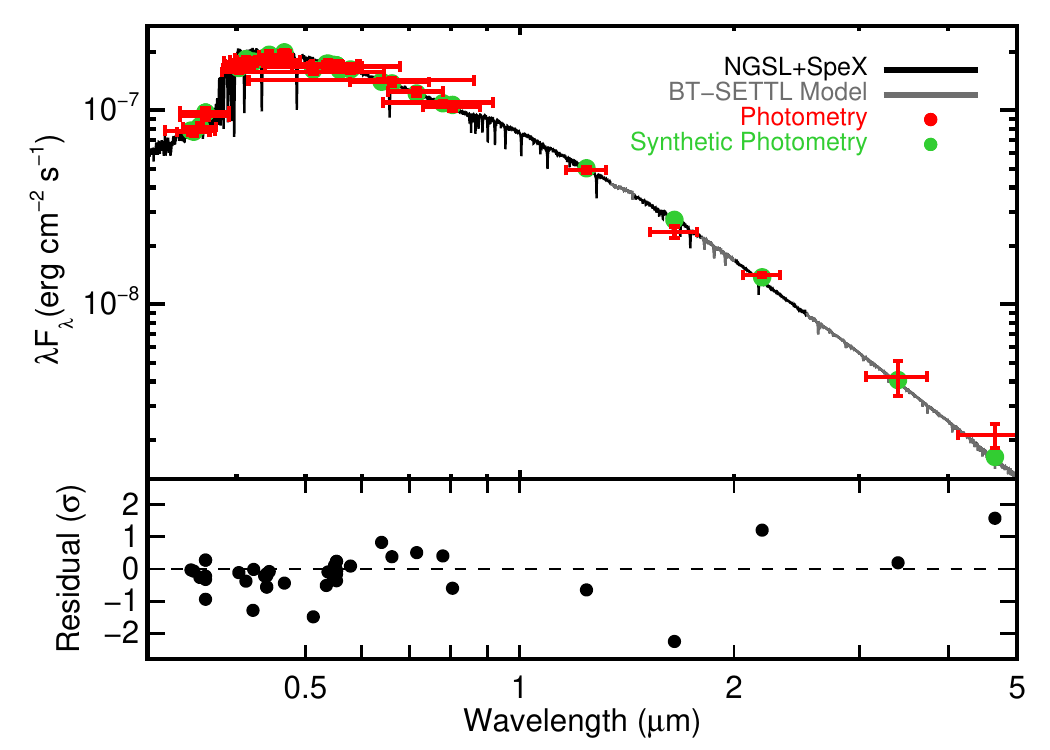}
		\caption{Spectral energy distribution and spectrum of 51 Eridani. The black line is the NGSL and SpeX spectrum, with grey regions indicating a model. Photometry is shown in red, with horizontal error bars indicating the width of the filter, while vertical show the photometric uncertainty. Synthetic photometry from the spectrum is shown in green. The bottom panel shows the difference between the spectrum and photometry in units of standard deviations.}
		\label{fig:sed}
	\end{figure}

	To estimate the bolometric flux ($F_{\rm bol}$) of \one, we use the spectral-energy distribution using available photometry from Gaia \citep{GAIA,Lindegren21}, the Two-Micron All-Sky Survey \citep[2MASS;][]{Skrutskie2006}, Tycho-2 \citep{Hog2000}, the Wide-field Infrared Survey Explorer \citep[WISE;][]{allwise}, and the General Catalog of Photometric Data \citep[GCPD;][]{Mermilliod1997}. We combine the spectrum of 51 Eridani from the STIS Next Generation Spectral Library \citep[NGSL;][]{Heap2007} (covering the near UV and optical) with one from Cool Stars library \citep{Rayner2009} (covering the near-infrared), filling any additional gaps with a stellar atmosphere \citep{Allard2011}. We scaled the combined spectrum to match the photometry, yielding an absolutely calibrated spectrum. The final fit is shown in Figure~\ref{fig:sed} and our bolometric flux measurement is in Table~\ref{tab:Vals}. More details on this fitting procedure, including our treatment of systematic uncertainties in the photometry and spectra are given in \citet{Mann2015b} and \citet{Mann2016b}.
	
	From the angular diameter and $F_{\rm bol}$, the stellar effective temperature, $T_{\rm eff}$, is calculated using the following form of the Stefan-Boltzmann equation:
	\begin{equation} \label{eq:teff}
		T_{\rm eff} = 2341 \cdot \Biggr(\frac{F_{\rm bol}}{\theta^{2}} \Biggl)^{1/4} K 
	\end{equation}
	\noindent In Equation~\ref{eq:teff}, the bolometric flux is in units of $10^{-8} \rm ~erg~s^{-1} cm^{-2}$ and $\theta$ is in units of mas. In the first initial calculation of $T_{\rm eff}$, $\theta_{UD}$ refers to the uniform disk angular diameter. In order to account for limb darkening, we use the Limb Darkening Coefficients table \citep{ClaretLDC} to solve for a linear $R$-band LDC with initial guesses of the star's $T_{\rm eff}$, $\log g$, and [Fe/H]. This value is used in a subsequent $\theta_{\rm LD}$ fit (Equation~\ref{eq:vsquare}) and a new temperature is derived using Equation~\ref{eq:teff} where $\theta$ is now referring to the limb-darkened angular diameter, $\theta_{\rm LD}$. We iterate on this process twice when no changes are seen. We determine the error for $\theta_{\rm LD}$ sampling the angular diameter in a Monte Carlo simulation, varying on a normal distribution of the uncertainties for wavelength ($\pm5$~nm), calibrator diameters ($5\%$ of the diameter), and limb-darkening coefficients ($\pm 0.02$). The resulting limb-darkened angular diameter is $\theta_{\rm LD} = 0.450 \pm 0.004$~mas.  
	
	Systematic variations (i.e. due to atmospheric effects or instrument alignments) can occur between nights. To investigate any systematic variations, we solve for the angular diameter using the data for each night separately, and found that they all agreed with $3\sigma$ of each other. We took a weighted average of the angular diameters from each individual night and found $\theta_{\rm LD} = 0.452 \pm 0.004$~mas, which agrees with the diameter from the combined fit within $1\sigma$. The weights for this average are $\frac{1}{\sigma^{2}}$ where $\sigma$ is the associated uncertainty for each data point. For the final limb-darkened angular diameter, we add the uncertainties from the combined diameter fit and the weighted average diameter in quadrature which can be found in Table~\ref{tab:Vals}.
	
\begin{figure}
		\includegraphics[width=0.95\columnwidth]{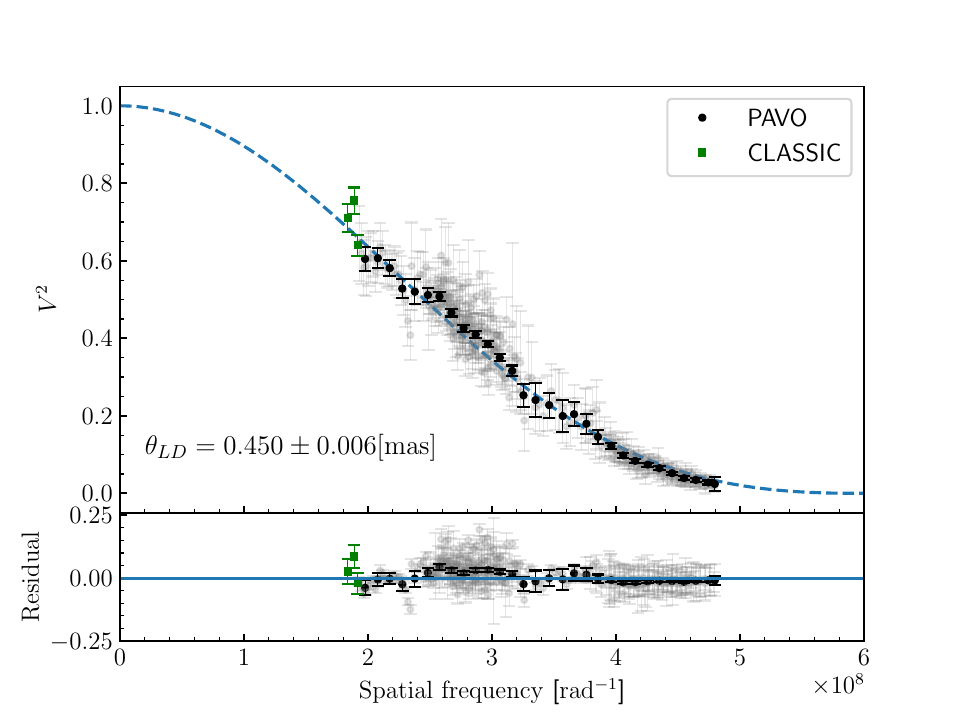}
		\caption{The top plot shows the calibrated squared visibilities and uncertainties from PAVO (in grey circles) and CLASSIC (in green squares). The PAVO data are binned in equal spacing and the weighted average is shown (in black circles). The dashed blue line is the visibility model (Eq.~\ref{eq:vsquare}) with the limb darkened coefficient of $\mu = 0.4516 \pm 0.0200$. The bottom plot displays the residuals from PAVO (unbinned in grey, binned in black) and CLASSIC (in green).The limb-darkened angular diameter is $0.450 \pm 0.006$~mas.}
		\label{fig:model}
	\end{figure}

	Figure~\ref{fig:model} displays the PAVO binned and calibrated visibilities, the CLASSIC data, and the final fit for angular diameter. The CLASSIC data presented are not used in the final fit for angular diameter because the observations do not adhere to the recommended observing minimum requirements of 2 nights, 2 calibrators, 2 baselines \citep{Boyajian12}. The data shown in Figure~\ref{fig:model} are there to demonstrate where CLASSIC lies on the curve (see Section~\ref{s:Disc} for further discussion.)
	
	\noindent A stellar luminosity, $L_{\star}$ is calculated using Equation~\ref{eq:lum}
	\begin{equation} \label{eq:lum}
		L_{\star} = 4\pi D^{2} F_{\rm bol}
	\end{equation}
	\noindent where $D$ is the distance (seen in Table \ref{tab:Vals}) and $F_{\rm bol}$ is the bolometric flux (seen also in Table \ref{tab:Vals}). The distance, $D$, is calculated using the zero-point corrected parallax, $33.44\pm 0.08$~mas (\citet{GAIA}, \cite{Lindegren21}). This yields a distance of $D = 29.93 \pm 0.07$~pc. The linear radius is then calculated using this new distance and Equation~\ref{eq:rad} below: 
	
	\begin{equation} \label{eq:rad}
		R_{\star} = \frac{\theta_{\rm LD}~D}{2}
	\end{equation}
	\noindent where $\theta_{\rm LD}$ is the limb-darkened angular diameter and $D$ is the zero-point corrected distance. The final parameters are presented in Table~\ref{tab:Vals}.
	\begin{table}[h!]
		\begin{center}
			\caption{Stellar and planetary parameters for \one~and 51 Eri b. Each property is listed in the left column, the corresponding value in the center column, and the source for each property in the right column. We use $\rm R_{\odot} = 6.957 \times 10^{8}$~m and $\rm L_{\odot} = 3.846 \times 10^{33} \rm ~erg~s^{-1}$.}
			\begin{tabular}{ccc} 
				\hline
				Property & Value & Source\\
				\hline
				$\theta_{\rm LD}~[\rm mas]$  & $0.450 \pm 0.006$       & Section ~\ref{S:direct}\\ 
				$F_{\rm bol}~[\rm 10^{-8}~erg~s^{-1}~cm^{-2}]$  & $20.5 \pm 0.3$  & Section ~\ref{S:direct} \\
				$\rm Parallax~[\rm mas]$  & $33.44\pm 0.08$      &  \cite{GAIA}, \\
				& &\cite{Lindegren21}\\
				$\rm Distance~[pc] $  & $29.93 \pm 0.07$     &   \cite{GAIA},\\ 
				& & \cite{Lindegren21} \\
				$ \rm [Fe/H]~[dex] $   & $0.13_{-0.02}^{+0.03}$        &  \cite{SwastikC2021}\\
				log $g$ $[\rm cm~s^{-2}]$ & $(4.07 - 4.10) \pm 0.21$ & \cite{Aren19}\\
				$T_{\rm eff}~[\rm K]$  & $7422 \pm 58 $       &  Section ~\ref{S:direct}\\
				$L_{\star}~[\rm L_{\odot}]$  & $5.72 \pm 0.096$      &  Section ~\ref{S:direct}\\
				$R_{\star}~[\rm R_{\odot}]$ & $1.45 \pm 0.02$    &  Section ~\ref{S:direct}  \\
				$\rm Age~[Myr]$ & $23.2^{+1.7}_{-2.0}$ & Section~\ref{s:agemass} \\
				$\rm Mass~[\rm M_{\odot}]$ & $1.550^{+0.006}_{-0.005}$ & Section~\ref{s:agemass} \\
				$\rm Planet$ $T_{\rm eff}~[\rm K]$ & $807 \pm 45$ & \cite{Brown2023} \\
				$ \rm Planet~Mass~[M_{\rm Jup}]$ & $4.1 \pm 0.4$ & Section~\ref{s:agemass}\\
				\hline
			\end{tabular}
			
			\label{tab:Vals}
		\end{center}
	\end{table}
	
	\section{Modeled Stellar and Planetary Properties}\label{s:empire}
	Using our measured effective temperature, luminosity, and radius, we estimate the age and mass of the star using stellar evolutionary models. Improved age estimates for the star means the age of the planet, 51 Eri b, is also improved given that the system is coeval. We then estimate the planet's mass through evolutionary modeling and provide further insight on how the planet formed. 
	
	\one~falls in a unique spot in the HR diagram. Previous works (summarized in the Introduction of \citealt{mama14}) indicated that the $\beta PMG$ consisted of pre-main sequence (PMS) stars. \cite{mama14} analyzed the kinematics of the $\beta PMG$ and concluded that the majority of the A0-F0 stars are either near or on the zero-age main sequence (ZAMS), which includes \one. 
	
	\subsection{Age and Mass Using Isochronal Modeling} \label{s:agemass}
	To estimate the age and mass of \one, we use stellar evolution models: the PAdova and TRieste Stellar Evolution Code (PARSEC) \citep{Bressan_PARSEC} and the Garching Stellar Evolution Code (GARSTEC) \citep{Weiss2008}. 
	
\begin{figure}
		\includegraphics[width=0.95\columnwidth]{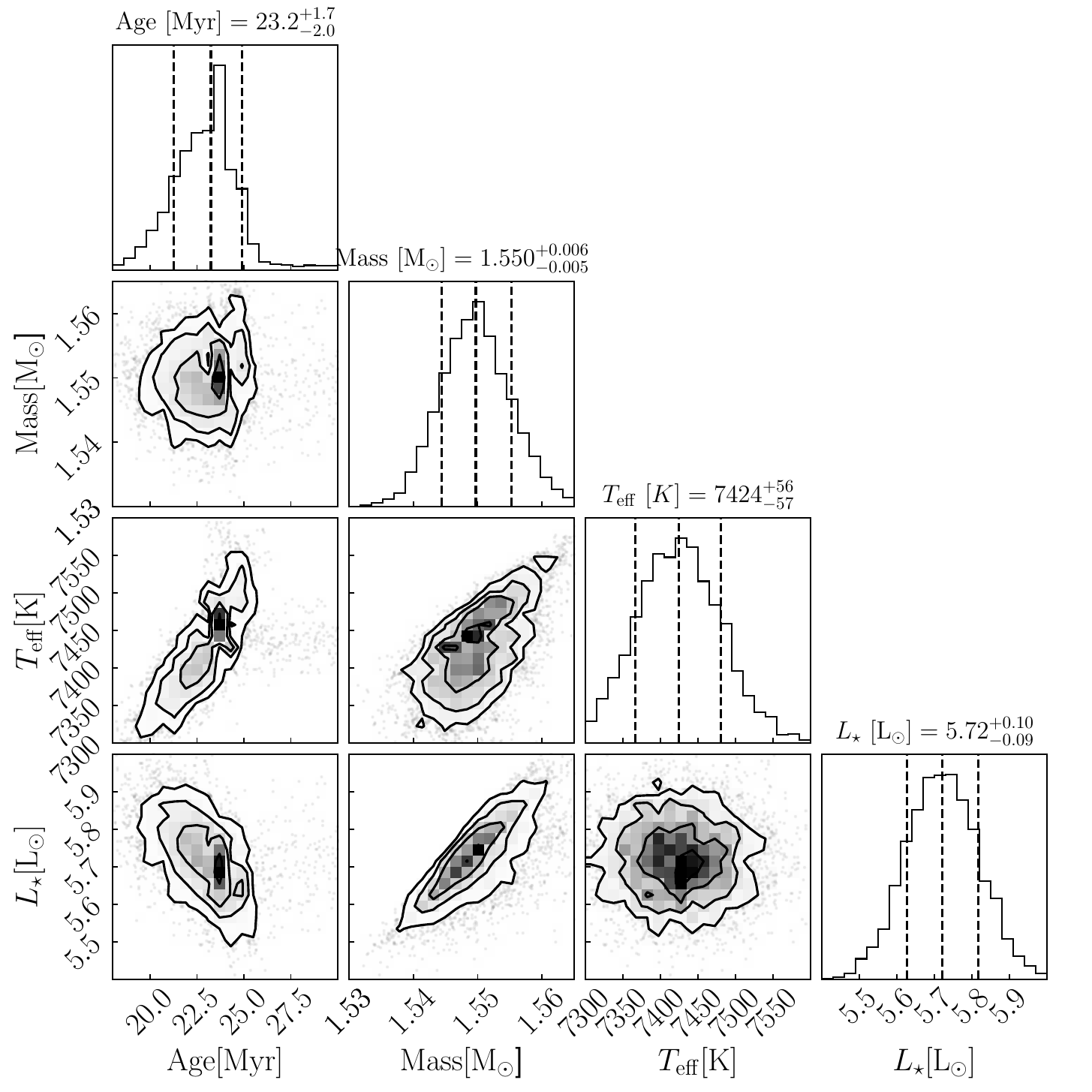}
		\caption{Corner plot for the PARSEC model displaying the results of a  Monte Carlo simulation of 5000 iterations to estimate the age and mass of \one. The dashed vertical lines indicate the quantiles for each histogram: $16 \%$, $50\%$, and $84\%$ from left to right. The left-most line corresponds to the lower bound uncertainty. The middle line corresponds to the accepted value. The right-most line corresponds to the upper-bound uncertainty. The solutions are as follows: an age of $23.2^{+1.7}_{-2.0}$ $\rm Myr$ and a mass of $1.550^{+0.006}_{-0.005}$ \msun.}
		\label{fig:PARSEC}
	\end{figure}

	We use the PARSEC version 1.2S model\footnote{Available at http://stev.oapd.inaf.it/cgi-bin/cmd} developed by \citet{Bressan_PARSEC} to create stellar isochrones given stellar priors. We generated isochrones for a range of ages between 0 and 50 Myr in steps of 0.5 Myrs in log space using a metal fraction of $Z = 0.0165$ (Table \ref{tab:Vals}). We interpolated the outputted isochrones to obtain a finer grid of points along each isochrone track. We then performed two separate two-dimensional interpolations over $\log (L/$\lsun) and $T_{\rm eff}$; one to extract an age and the other to extract a mass. We use a Monte Carlo simulation to obtain the errors for the age and mass estimates by sampling the $T_{\rm eff}$ and luminosity in a normal distribution. The posterior distributions are shown in Figure \ref{fig:PARSEC}. The final solution results in an age of $23.2^{+1.7}_{-2.0}$ Myr and a mass of $1.550^{+0.006}_{-0.005}$ \msun. 
	
\begin{figure}
		\includegraphics[width=0.95\columnwidth]{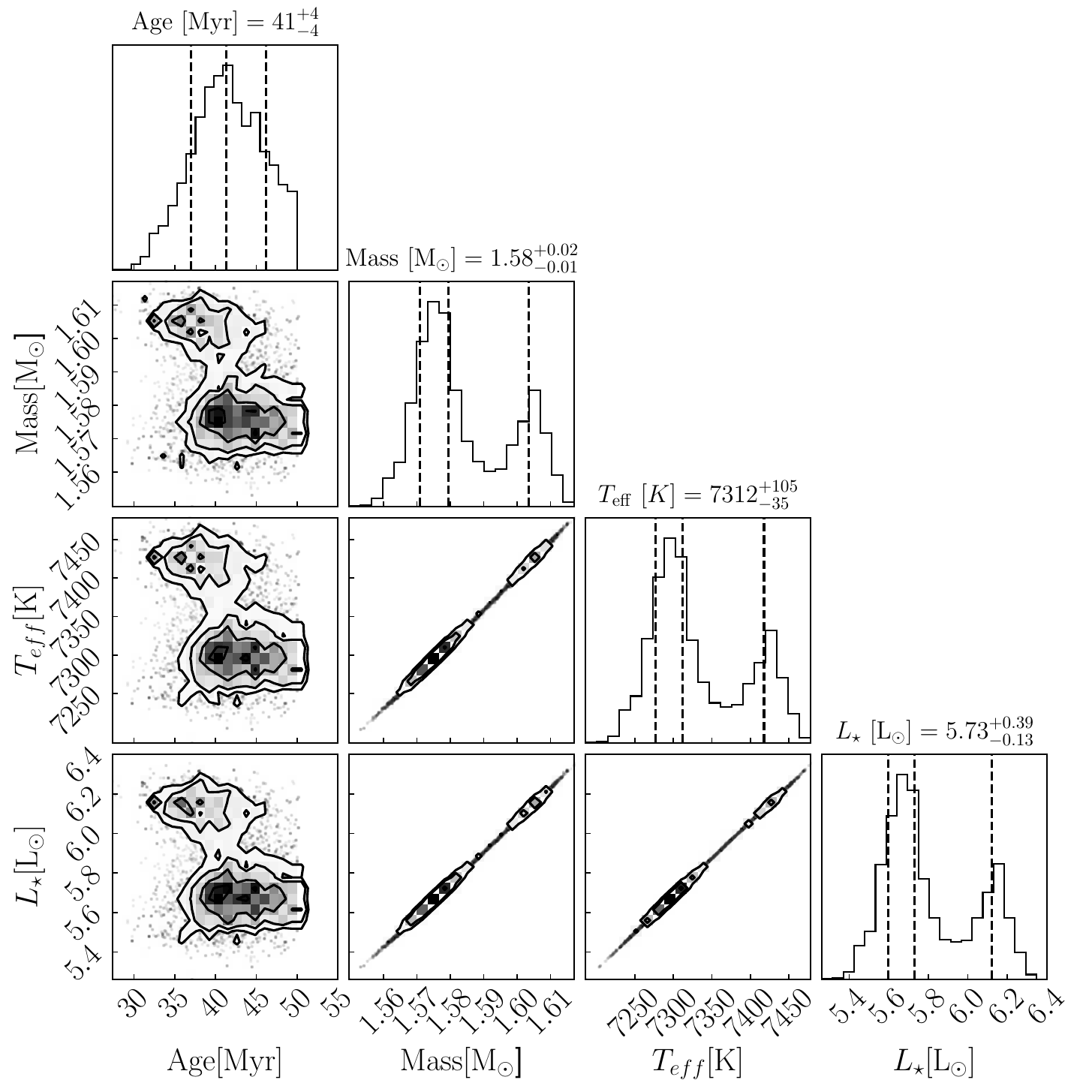}
		\caption{The above plot displays the corner plot for the GARSTEC model run through {\fontfamily{qcr}\selectfont bagemass}. The dashed vertical lines indicate the quantiles for each histogram: $16 \%$, $50\%$, and $84\%$ from left to right. The left-most line corresponds to the lower bound uncertainty. The middle line corresponds to the accepted value. The right-most line corresponds to the upper-bound uncertainty. The solutions are as follows: an age of $41\pm4$ Myr and a mass of $1.58_{-0.01}^{+0.02}$ \msun.}
		\label{fig:GARSTEC}
	\end{figure}

	We then used the GARSTEC models \citep{Weiss2008} through the implementation of {\fontfamily{qcr}\selectfont bagemass}. {\fontfamily{qcr}\selectfont bagemass} \citep{MaxBAGE} is a program written in Fortran that uses stellar evolution models to estimate the mass and age of a star and produces posterior probability distributions calculated using Bayesian methods. The GARSTEC model chosen has an $\alpha$ mixing length of 1.78 and a He-enhancement value of 0. The priors used are $T_{\rm eff}$, $\log (L/$\lsun), and [Fe/H] (Table \ref{tab:Vals}). We fixed the metallicity and the age is given a range between 0 and 50 Myr. The posterior distributions are shown in Figure \ref{fig:GARSTEC}. The final solution determined gives an age of $41\pm 4$ Myr and a mass of $1.58_{-0.01}^{+0.02}$ \msun. 
	
	The age and mass estimates determined by the PARSEC model and the GARSTEC model are vastly different, with the GARSTEC model estimating an age almost twice that of the PARSEC model. The GARSTEC model could not return the prior distributions for $T_{\rm eff}$ and log($L/$\lsun) within the uncertainties or stay within the range of [Fe/H] priors given. Looking at the posterior distributions, we see that the GARSTEC model returns a bimodal solution and skews the priors, reflecting the model not returning the priors sufficiently well enough. This bimodal distribution could be due to the star's location on the HR diagram. Although both models incorporate the PMS phase and the ZAMS phase of stellar evolution, the position of this star being right at the transition point could cause each model to either favor the pre-main sequence (PMS) side or the early zero-age main sequence (ZAMS) side of this transition point. However, the PARSEC model is able to return the priors well and produce a single mode solution. We adopt for the final results of this work, the PARSEC age and mass of $23.2^{+1.7}_{-2.0}$ Myr and $1.550^{+0.006}_{-0.005}$ \msun.
	
	The determination of age and mass is reliant on the priors of effective temperature and luminosity, so it is logical that the uncertainties of age and mass would be in turn reliant on these priors. In addition to the effect from temperature and luminosity, the uncertainties for age and mass are reliant on the choice in model chosen to determine these parameters \citep{Tayar2022}. The age is more affected by the choice in model, especially for stars closer to the ZAMS or PMS, where \cite{Tayar2022} describes the differences in age being near $100\%$ in these regions of parameter space. The differences between the PARSEC and the GARSTEC models used to estimate the age of 51 Eridani is another example of the discrepancy in age between models.
	
	\subsection{Analysis of the Planetary Companion 51 Eri b}\label{s:Planet}
	
	The next step in the analysis of 51 Eri b is to estimate a mass using the assumption that the planet is the same age as the star (see Section~\ref{s:agemass}). We use the Sonora Bobcat models \citep{Marley2021} which were developed to study L-, Y-, and T- type brown dwarfs and self-luminous exoplanets. 51 Eri b is considered to be a self-luminous planet but it has been debated on whether 51 Eri b is an L- or T- type brown dwarf given its unique strong methane absorption features that are also seen in T- type brown dwarfs and similar near-IR colors to L-type brown dwarfs \citep{maci15}. 
	
\begin{figure}
		\includegraphics[width=1.0 \columnwidth]{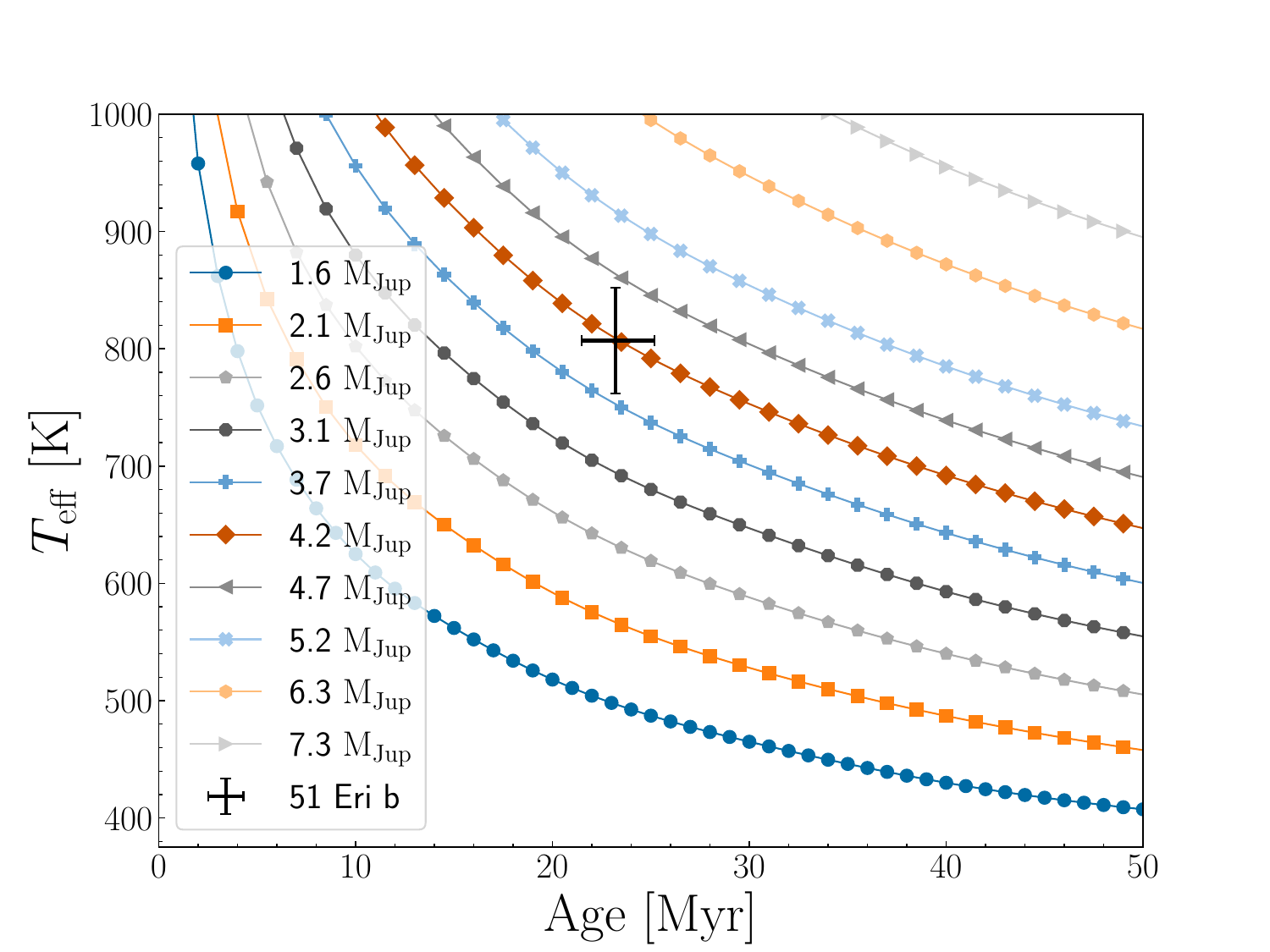}
		\caption{The planet iso-mass models from \citet{Marley2021}. Iso-mass models were created using the The Sonora Bobcat model evolutionary tables that hold mass fixed. A one-dimensional interpolation was done to create a finer grid of points for each line. Then, a two-dimensional interpolation was done over $T_{\rm eff}$ and age to determine a planetary mass. For results, see Figure~\ref{fig:planetcorner} and Table~\ref{tab:Vals}.}
		\label{fig:planetcurves}
	\end{figure}

\begin{figure}
		\includegraphics[width=0.95\columnwidth]{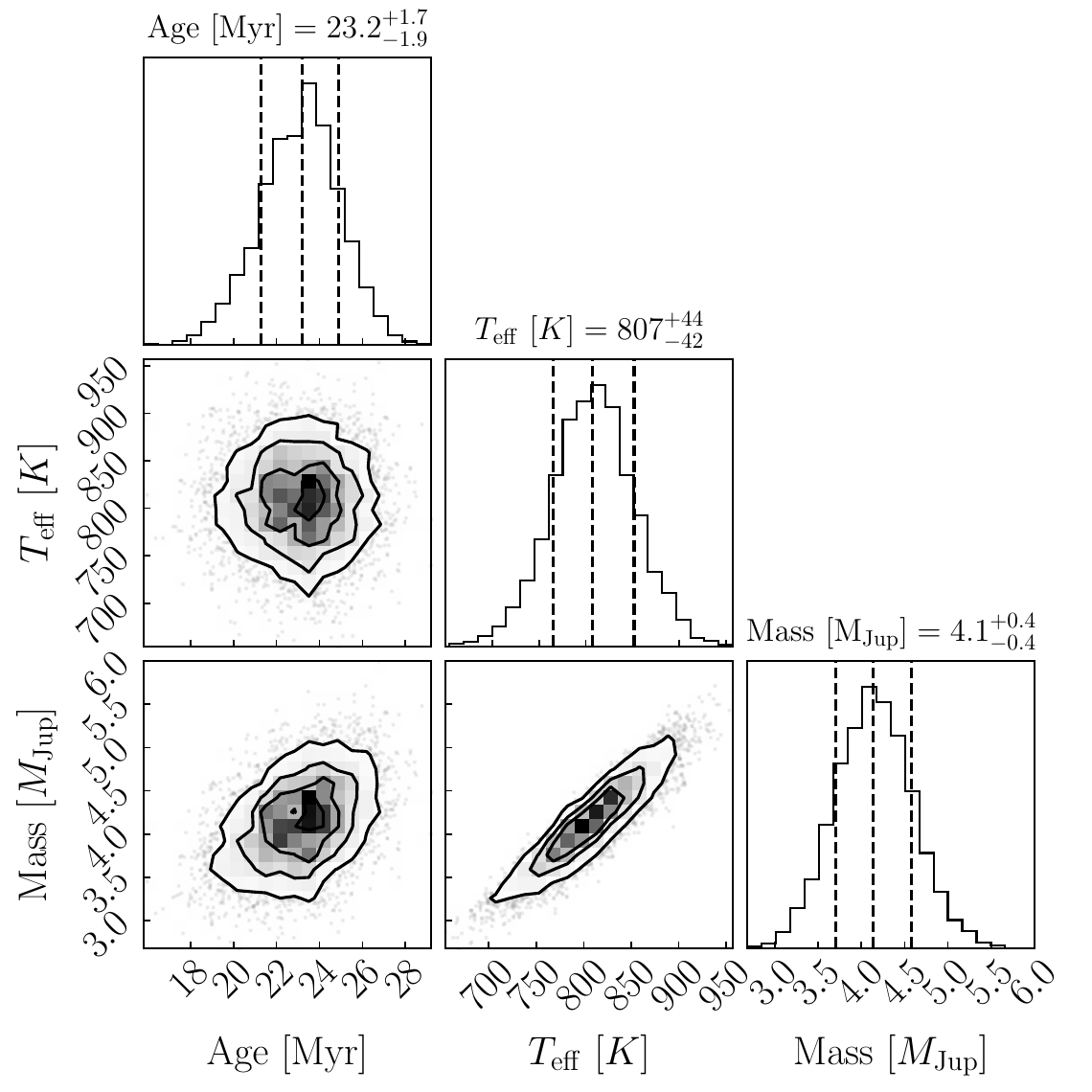}
		\caption{Corner plot for the Sonora Bobcat planet iso-mass model displaying the restuls of a Monte Carlo simulation of 5000 iterations to estimate the mass of 51 Eri b. The dashed vertical lines indicate the quantiles for each histogram: $16 \%$, $50\%$, and $84\%$ from left to right. The left-most line corresponds to the lower bound uncertainty. The middle line corresponds to the accepted value. The right-most line corresponds to the upper-bound uncertainty. The final results of this simulation give a mass of $4.1\pm 0.4$~\mjup. See Section~\ref{s:Planet} for details. }
		\label{fig:planetcorner}
	\end{figure}

	The Sonora Bobcat models produce evolutionary tables that hold either age, mass, or bolometric luminosity fixed. For this paper's purpose, we use the model holding mass fixed. In our analysis, we use the planet's effective temperature of $T_{\rm eff} = 807 \pm 45 $ K \citep{Brown2023}. Figure~\ref{fig:planetcurves} shows the Sonora Bobcat models as age vs $T_{\rm eff}$ along with the position of 51 Eri b. To determine a mass, we first perform a one-dimensional interpolation to create a finer grid of points for each of the iso-mass lines. We then perform an additional two-dimensional interpolation over the effective temperature and age and find the corresponding mass. To obtain uncertainties for the mass, we ran a Monte Carlo simulation, sampling the age and $T_{\rm eff}$ on a normal distribution. For the age, an asymmetrical normal distribution was created to account for the uneven uncertainties. Figure \ref{fig:planetcorner} shows the posterior distribution for this simulation. The final mass of 51 Eri b is $4.1\pm 0.4$~\mjup. 
	
	\section{Discussion} \label{s:Disc}
	\subsection{Stellar Parameters}
	We measure a limb-darkened angular diameter for \one~of $\theta_{\rm LD} = 0.450 \pm 0.006$ mas (Section~\ref{s:empire}). We also measure an angular diameter of $\theta_{LD} = 0.425 \pm 0.026$ mas using a limb-darkening coefficient of $\mu = 0.2582$ from our CLASSIC data alone, which is consistent with the PAVO diameter within $1\sigma$. The PAVO measurement has a precision of less than $1\%$, compared to the CLASSIC measurement which has a precision of just over $6\%$. This is due to the greater spatial frequency coverage obtained with PAVO’s multiple wavelengths channels, as well as sampling further down the visibility curve to higher spatial frequencies. More coverage, especially at higher spatial frequencies, ensures that the visibility squared fit is more accurate. \cite{Simon2011} measure an angular diameter of \one~also using CHARA's CLASSIC beam combiner. Their measurements were taken in both $H$- and $K^{'}$-band on the E1/W1 baseline (314 m). \cite{Simon2011}'s angular diameter of $\theta_{\rm LD} = 0.518 \pm 0.009$ mas is over $7\sigma$ off from our PAVO angular diameter. We suspect that this disagreement is due to the observation strategy used by \cite{Simon2011} who observed on a single baseline (E1/W1) and only used one calibrator in their analysis. This observation strategy is dangerously insensitive to verifying whether or not the single calibrator they used is good. In contrast, our measurements were taken using multiple calibrators over multiple nights (see Table~\ref{tab:Obs}). The angular diameter of \one~is near the resolution limit ($\frac{\lambda}{2B}$) of the CHARA Array on the longest baselines in the near-infrared. Analysis in the literature shows that interferometric measurements of under-resolved stars tend to systematically over-estimate their angular diameters (\cite{Casagrande2014}, \cite{White2018}, \cite{Tayar2022}). However, our observations using CLASSIC data alone do not support this trend as they are in agreement within $1\sigma$ of the PAVO observations. 
	
	\cite{Simon2011} also performed an analysis to estimate the age and mass of \one. This analysis used the PMS evolution models developed by \cite{Seiss2000} and the Yonsei-Yale (Y2) models \citep{Y2}. \cite{Simon2011} compared their absolute angular diameter (angular diameter scaled to a distance at 10 pc) with those predicted by both models and found an average age of $13\pm2$ Myr and a mass of $1.75\pm0.05$~\msun. Our age is almost twice that of \cite{Simon2011} and our mass is $4\sigma$ off of \cite{Simon2011}'s. If we scale our angular measurement to a distance of 10 pc, it would fall closer to the older isochrones in Figure 3 in \cite{Simon2011} for the same $V-K$, suggesting that the difference in angular diameter seems to be the main source of discrepancy. Additional errors are subject to arise in transforming the theoretical properties calculated by the models to observational quantities, such as magnitudes and color indexes from color tables. Comparing our directly measured $T_{\rm eff}$ and luminosity to the model predictions ensures that any ambiguous errors that result from the color tables will not affect our results.

	\cite{Simon2011} also used a $V-K$ vs. $M_{K}$ diagram to
	estimate age and mass of~\one. The mass results were around $0.2$~\msun smaller than their initial analysis of comparing the model-predicted angular diameters and yielded older ages (around 5 Myr), creating a noted discrepancy between the two methods for age and mass. However, the $V-K$ vs. $M_{K}$ result estimates align with our results.
	
	The age we determine with the PARSEC model ($23.2^{+1.7}_{-2.0}$ Myr) agrees with the age estimates found in the literature of both the star itself and that of the $\beta PMG$. These age estimates have been determined through a variety of methods from isochronal ages to lithium depletion methods. \cite{mama14} conclude with an isochronal age of $22\pm 3$ Myr, where their value compares favorably to several lithium depletion boundary ages from \cite{Binks} ($21\pm4$ Myr) and \cite{Malo} ($26\pm3$ Myr). 
	
	A more recent study on the $\beta PMG$ from \cite{Mir20} concludes with an age estimate of $18.5_{-2.4}^{+2.0}$ Myr through dynamical age trace-back methods using precise Gaia DR2 astrometry and ground based radial velocities. \cite{Couture23} developed a method to correct the trace-back age method that reduces the systemic errors from a combination of uncorrected gravitational red-shift and convective blue-shift absolute radial velocity measurements and the random errors from the parallax, proper motion, and additional radial velocity measurements. Their final resulting age is $20.4 \pm 2.5$ Myr for the $\beta PMG$. Our age estimate compares favorably to \cite{Couture23}'s dynamical age estimates of the $\beta PMG$.

	\subsection{A quick look at available TESS data}
	The Transiting Exoplanet Survey Satellite (TESS: \cite{TESS}) observed \one~ in two sectors: Sector 5 and Sector 32. \cite{Sepul2022} investigated the variability seen in the TESS light curve data and concluded that \one~is a $\gamma$-Dor pulsator. We downloaded the available 2 minute cadence data using {\fontfamily{qcr}\selectfont lightkurve} \citep{lightkurve}. The data were flattened and binned. From these data, we constructed a periodogram of the processed TESS data to search for significant exoplanet signals. We extracted significant frequencies and tested to see if a possible planet signal could be found. There were no significant possible planet signals. The light curves did show significant variability which does corroborate the findings of \cite{Sepul2022} in \one~being a $\gamma$-Dor pulsator. 
	
	\subsection{Planetary Parameters}
	We derive a mass for 51 Eri b to be $4.1\pm 0.4$~\mjup~through the Sonora Bobcat evolutionary models (Section~\ref{s:Planet}).  Other studies have estimated the mass of 51 Eri b through a variety of methods, such as atmospheric modeling and evolutionary modeling. Several works present exploration of both methods, i.e. using atmospheric modeling to obtain planetary parameters such as effective temperature, surface gravity, luminosity, and radius, and using these parameters along with an adopted age in evolutionary models to estimate a mass. 
	
	The initial discovery announcement of 51 Eri b \citep{maci15} performed atmospheric modeling to estimate the planet parameters, such as effective temperature and luminosity. \cite{maci15} used two separate models, a cloud-free and a partial cloudy model to obtain these parameters. The cloud-free model resulted in a $T_{\rm eff} = 750$~K and a log($L/\rm L_{\odot}) = -5.8$. The partial cloudy model gave a $T_{\rm eff} = 700$~K and a log($L/\rm L_{\odot}) = -5.6$. \cite{maci15} determined that the luminosity derived from the models does not change much with either model and used the luminosity along with age to estimate a mass based off of formation models. Given hot-start formation, the mass of 51 Eri b is $\sim 2$ \mjup. With a cold-start formation, the mass falls in a range of 2-12 \mjup. Our mass of $4.1$~\mjup~is well within the cold-start formation range of ages but is over double that of the hot-start formation age.

	\cite{Sam2017} presented the first spectro-photometric measurements in the $Y$- and $K$- bands of 51 Eri b using VLT/SPHERE. This study used atmospheric modeling using {\fontfamily{qcr}\selectfont petitCODE} - Cloudy to obtain planetary parameters. The resulting temperature, $T_{\rm eff} = 760 \pm 20$~K, radius of $1.11^{+0.16}_{-0.14}$~\rjup, and surface gravity of $4.26 \pm 0.25$ (cgs-units) are used in a variety of methods to obtain mass estimates. \cite{Sam2017} used surface gravity and radius relations to get a mass of $9.1^{+4.9}_{-3.3}$~\mjup~which is \cite{Sam2017} also used radius and effective temperature relations to derive a luminosity and a planet formation model which relates luminosity and mass. The range in luminosity rules out the cold-start formation model and resulted in masses between 2.4 and 5~\mjup~for the hot-start model and a larger spread of masses between 2 and 12~\mjup~for the warm-start model. Our mass also agrees with both mass ranges for the hot-start and warm-start models. 
	
	\cite{dup22} performed a cross calibration of Hipparcos-GAIA astrometry and the orbit fitting code {\fontfamily{qcr}\selectfont orvara}. Their results provided orbital parameters as well as an upper limit on the mass of 51 Eri b of $< 11$~\mjup. Through private communication, another run of the {\fontfamily{qcr}\selectfont orvara} fit was made using the new age and mass of \one~presented in this paper. The new upper limit $2 \sigma$ constraint on the mass of 51 Eri b is 9.5 \mjup. Our mass of $4.1$~\mjup~clearly falls under this upper limit. \cite{dup22}'s result also indicated that the cold-start formation is ruled out given the derived initial entropy. They suggested that 51 Eri b formed similarly to other directly imaged planets which indicate either the warm or hot-start scenarios. 
	
	The most recent study of 51 Eri b, done by \cite{Brown2023}, revisited the work by \cite{Sam2017} and obtained new observations of 51 Eri b with VLT/SPHERE at a higher S/N than \cite{Sam2017}'s original data. \cite{Brown2023} used a new atmospheric retrieval code, {\fontfamily{qcr}\selectfont petitRADTRANS}, to update the $T_{\rm eff}$ and the resulting posterior distributions for the log$g$ and planet radius to obtain a mass of $3.9\pm 0.4$~\mjup. \cite{Brown2023} labeled this result as the nominal model. \cite{Brown2023} also performed an analysis using evolutionary models from \citet{Baraffemodels} with two separate ages: 10 Myr \citep{Lee2022} and 20 Myr \citep{maci15}, where they found masses of $2.4$~\mjup~and $2.6$~\mjup~respectively. The mass presented in this work ($4.1\pm 0.4$~\mjup) agrees well with the nominal result obtained by \cite{Brown2023}. 
	
	Our mass falls within the ranges of both the hot-start and the warm-start models as described by both \cite{Sam2017} and \cite{Brown2023}. With our analysis, we can rule out a purely core accretion model as the main formation mechanism. Either scenario is likely but future observations and analyses are needed to further test each model and develop a more conclusive argument. 
	
	\section{Conclusion}
	In this work, we present interferometric observations of the directly imaged exoplanet host star \one~taken with both the PAVO and CLASSIC beam combiners at the CHARA Array. These observations resulted in a highly precise angular diameter measurement of \one, $\theta_{\rm LD} = 0.450\pm0.006$~mas. From this angular diameter, we calculate the effective temperature, $T_{\rm eff} = 7422 \pm 58$~K, luminosity, $L_{\star} = 5.7\pm0.1$~L$_{\odot}$, and linear radius, $R_{\star} = 1.45\pm 0.02$~\rsun. We use the PARSEC isochrones to derive an age of $23.2^{+1.7}_{-2.0}$~Myr and mass of $1.550^{+0.006}_{-0.005}$~\msun~of \one. For 51 Eri b, we estimate a mass of is $4.1\pm 0.4$~\mjup~using the Sonora Bobcat evolutionary models. 
	
	Future analysis is encouraged using the JWST data taken in 2022 with NIRCAM and MIRI. Additional spectra can provide more constraints on planet composition, effective temperature, and luminosity when used with the atmospheric retrievals developed for JWST. \cite{Brown2023} suggested the use of the Mid-Infrared ELT Imager and Spectrograph \citep{Quanz2015} which will increase the wavelength coverage up to 13$\mu m$. 
	
	Observations of other members of the $\beta PMG$ can help further the age constraints on these objects. In Table A1 in \cite{AlonsoF15}'s paper, they list 185 members and candidate members of the $\beta PMG$. At least two F type stars are observable with PAVO at the CHARA Array and with future instruments being commissioned, such as the SPICA instrument (\cite{SPICA}, \cite{Mourard2022}), the increased sensitivity will allow for more targets in this group to be observed. In addition, there are four stars that are accessible in the southern sky with VLTI, that have angular diameters ranging between 0.46 mas and 0.75 mas (estimated using the surface brightness relationships in \citealt{Adams2018}). Two of these stars are observable with the current instrumentation available at VLTI. With the planned improvements to VLTI, such as adding the BIFROST \citep{Bifrost} instrument, the other two stars could be observable in the near future. Observing and characterizing more stars within the $\beta$PMG will help to estimate the age of the group and the individual stars within, giving us more insight on stellar evolution.
	
	\section*{Acknowledgements}
	This work is based upon observations obtained with the Georgia State University Center for High Angular Resolution Astronomy Array at Mount Wilson Observatory. The CHARA Array is supported by the National Science Foundation under Grant No. AST-1636624 and AST-2034336. Institutional support has been provided from the GSU College of Arts and Sciences and the GSU Office of the Vice President for Research and Economic Development.
	
	CHARA telescope time was granted by NOIRLab through the Mid-Scale Innovations Program (MSIP). MSIP is funded by NSF.
	
	CHARA Array time was granted through the NOIRLab community-access program (NOIRLab Prop. ID: 2021A-0141; PI: T. Boyajian).
	
	CHARA Array time was granted through the NOIRLab community-access program (NOIRLab Prop. ID: 2021A-0247; PI: T. Ellis).
	
	We thank Dr. Gururaj A. Wagle for his assistance in understanding interpolation and Henry Ngo for the extremely helpful suggestions in debugging interpolation code. We also thank Nageeb Zaman and Dr. Jonas Kl\"uter for their assistance in developing an asymmetrical normal distribution for Monte Carlo simulations used in this work. 
	
	This research has made use of the VizieR catalogue access tool, CDS, Strasbourg, France (DOI :  10.26093/cds/vizier). The original description of the VizieR service was published in 2000, A\&AS 143, 23 \citep{vizier}. 
	
	This research has made use of the Jean-Marie Mariotti Center JSDC catalogue \footnote{Available at http://www.jmmc.fr/catalogue\_jsdc.html}.
	
	This research has made use of the Jean-Marie Mariotti Center \texttt{Aspro} \footnote{Available at http://www.jmmc.fr/aspro} service. 
	
	This paper includes data collected by the TESS mission. Funding for the TESS mission is provided by the NASA's Science Mission Directorate.
	
	This research made use of Lightkurve, a Python package for Kepler and TESS data analysis \citep{lightkurve}.
	\section*{Financial Support}
	A.E. and T.S.B. acknowledge support by the National Support Foundation under Grant No. AST-2205914.
	\section*{Conflicts of Interests}
	None
	\section*{Data Availability}
	
	All interferometric data are available in the CHARA archive\footnote{Available at https://www.chara.gsu.edu/observers/database}. The TESS observations are available in the Barbara A. Mikulski Archive for Space Telescopes (MAST) archive.

	\bibliographystyle{paslike}
	\bibliography{HD29391sources}
	
\end{document}